\documentclass{article}
\usepackage{spconf,amsmath,epsfig}

\usepackage{amsmath,amsfonts,amssymb,amsbsy} 

\usepackage{hyperref} 

\usepackage{cite} 
\usepackage{color}

\usepackage{colortbl}

\newcommand{\images}{./images/} 
\newcommand{\imagesdim}{./images/imres/}

\title{Cloud detection machine learning algorithms for PROBA-V}

\name{Luis G{\'o}mez-Chova, Gonzalo Mateo-Garc{\'i}a, Jordi Mu{\~n}oz-Mar{\'i}, Gustau Camps-Valls
\thanks{This work has been partially supported by 
the Spanish Ministry of Economy and Competitiveness (MINECO, TEC2016-77741-R and TIN2015-64210-R), 
the European Space Agency (ESA IDEAS+ research grant, CCN008), 
and ERC Consolidator Grant SEDAL ERC-2014-CoG 647423. 
\newline
Preprint corresponding to the paper published in 2017 IEEE International Geoscience and Remote Sensing Symposium (IGARSS), Fort Worth, TX, USA, pp. 2251-2254, DOI: 10.1109/IGARSS.2017.8127437.
}}
\address{Image Processing Laboratory (IPL), University of Valencia, Spain
}

\begin{document}

\maketitle

\begin{abstract}
This paper presents the development and implementation of a cloud detection algorithm for Proba-V. 
Accurate and automatic detection of clouds in satellite scenes is a key issue for a wide range of remote sensing applications. With no accurate cloud masking, undetected clouds are one of the most significant sources of error in both sea and land cover biophysical parameter retrieval. 
The objective of the algorithms presented in this paper is to detect clouds accurately providing a cloud flag per pixel. For this purpose, the method exploits the information of Proba-V using statistical machine learning techniques to identify the clouds present in Proba-V products. 
The effectiveness of the proposed method is successfully illustrated using a large number of real Proba-V images. 
\end{abstract}


\section{Introduction}\label{sec:intro}

The main objective of this work is to propose a cloud detection algorithm for Proba-V~\cite{Dierckx14}. Images acquired by Proba-V instrument, which works in the visible and infrared (VIS-IR) ranges of the electromagnetic spectrum, may be affected by the presence of clouds. 

\begin{figure}[]
\centering
\includegraphics[width=8.5cm]{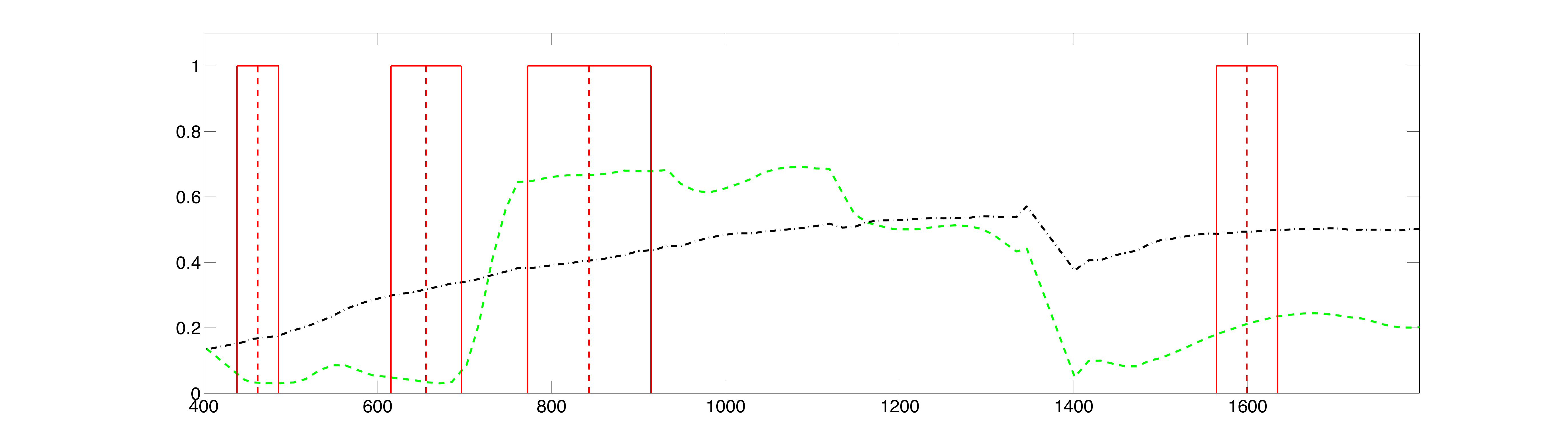} \\
\vspace{-0.1cm}{\footnotesize wavelength (nm)}
\caption{\small \label{fig:ProbaV_bands} Proba-V channels (boxes) superimposed to a reflectance spectra of healthy vegetation and bare soil (dash-dotted lines).} 
\end{figure}

Cloud detection approaches, also referred to as cloud masking, are generally based on the assumption that clouds present some useful characteristics for its identification. 
The simplest approach to cloud detection in a scene is the use of a set of static thresholds (e.g. over reflectance or temperature) applied to every pixel in the image, which provides a cloud flag (binary classification) \cite{Wang06}. 
However, the Proba-V instrument presents a limited number of spectral bands (Blue, Red, NIR and SWIR) which makes cloud detection particularly challenging since it does not present thermal channels or a dedicated cirrus band (Fig.~\ref{fig:ProbaV_bands}). 
On the one hand, thicker clouds should be easily detected and masked out from visible and near-infrared Proba-V bands, but this is not true for thiner clouds, which are semitransparent to solar radiation. Moreover, bright pixels, such as ice and snow in the surface, can be misclassified as clouds. Bright land covers and clouds have a similar reflectance behavior, thus thresholds on reflectance values do not solve the problem. 
On the other hand, signal coming from optically-thin semitransparent clouds is mostly affected by surface contribution, and it ranges from very low to extremely high values depending on whether the cloud is over water or ice, respectively. Therefore, they are extremely difficult to detect from reflectance properties in VNIR data.  
These problems preclude the use of simple approaches based on static thresholds and suggest the use of more advanced cloud masking methods \cite{GomezChovaTGARS07,GomezChovaIGARSS07}. 

Current Proba-V cloud detection uses multiple thresholds applied to the blue and the SWIR spectral bands~\cite{Lisens00}, but the definition of global thresholds is practically impossible. Hence, for next Proba-V reprocessing~\cite{Wolters15}, monthly composites of cloud-free reflectance in the blue band are used to define dynamic thresholds depending on the land cover type. 
Few works using more sophisticated machine learning tools have been presented so far in the cloud detection literature, such as Bayesian methods \cite{Murtagh03}, fuzzy logic \cite{Ghosh06}, artificial neural networks \cite{Torres-Arriaza03}, or recently kernel methods \cite{GomezChovaIGARSS07,GomezChovaTGARS10}.  

In this context, the European Space Agency (ESA) has started the study {\em `Clouds Detection Algorithms for Proba-V'} in order to propose and compare different cloud detection approaches for Proba-V. This paper presents our contribution in the framework of this Proba-V Clouds Detection Round Robin experiment. 
The proposed cloud detection algorithms rely on advanced non-linear methods capable of exploiting the information of Proba-V features in order to improve the cloud masking products.

\section{Methodology}\label{sec:method}

The cloud masking process relies on the extraction of meaningful physical features (e.g. brightness and whiteness) that are combined with spatial features to increase the cloud detection accuracy. 
Then, a supervised pixel-based classification, based on the TOA reflectance and on a manually labeled training set, is applied to these features providing the pixel label (\emph{cloud} or \emph{cloud free}). 
The supervised classifiers to be developed and tested on this data should allow the use of a high number of input features and allow an easy integration of heterogeneous sources of information.

\subsection{Proba-V Data and Ground Truth}\label{sect:data}  

In this work, we consider as input data Proba-V Level 2A products with TOA reflectance, i.e. the four Proba-V bands are radiometrically and geometrically corrected and resampled at 333m. 
The available data set for the Proba-V Round Robin\footnote{\href{https://earth.esa.int/web/sppa/activities/instrument-characterization-studies/pv-cdrr}{http://earth.esa.int/web/sppa/activities/instrument-characterization-studies/pv-cdrr}} exercise consists of 331 products acquired in four days covering the four seasons: 21/03/2014,  21/06/2014, 21/09/2014, and 21/12/2014. 
A reduced set of Proba-V data for a number of representative sites worldwide is used to train the algorithm and validate its performance. 

In order to train statistical machine learning models from real data, a representative number of  samples have to be labelled as \emph{cloud-contaminated} or \emph{cloud-free} samples. 
To label in a semi-automatic way a sufficient number of pixels from the Proba-V images, we have adapted the user-driven methodology proposed for MERIS in \cite{GomezChovaTGARS07}  to the Proba-V images, where the labeling of cloud clusters found in the image is done by the user.

\subsection{Feature Extraction} \label{sec:feature_extraction}

Several physically-inspired features can be extracted from the spectrum before applying the classification methods in order to improve their performance. 
In this work, we take advantage of previous research and, rather than working with the spectral reflectance only, physically-inspired features are extracted in order to increase the separability of clouds and surface covers \cite{Mei17}. 
The final set of features that are analyzed in the frame of this work are listed in Table~\ref{table:cloud_features}. 

Together with the spectral features, basic spatial features are extracted at different scales: the mean ($\mu$) and standard deviation ($\sigma$) are computed for each pixel-based feature at two different scales in $3\times3$ and $5\times5$ windows. 
Summarizing, we consider the four Proba-V spectral channels ($4$), the spectral features described in Table~\ref{table:cloud_features} ($10$), and the mean ($\mu$) and standard deviation at two different scales, which are computed for each pixel-based feature ($(4+10)\times4$). That results in a total number of $70$ possible input features ($4$ reflectance bands, $10$ spectral features, $56$ spatio-spectral features). 
Moreover, in order to reduce the complexity of the trained classifiers, we define two sets of features with the first 20 and 40 most relevant features that will be compared in the classification experiments.

\begin{table}[t]
\caption{ \small \label{table:cloud_features} Cloud features extracted from Proba-V images.}
\begin{center}
\footnotesize
\begin{tabular}{p{3cm}|p{4.5cm}}
\hline
\hline
\bf{Cloud Feature} & Feature \\
\hline
Brightness  & $x_{Br}$ \\
Brightness VIS & $x_{Br,VIS}$ \\
Brightness NIR & $x_{Br,NIR}$ \\
Whiteness & $x_{Wh}$ \\
Whiteness VIS & $x_{Wh,VIS}$ \\
Whiteness NIR & $x_{Wh,NIR}$ \\
Snow NDSI & $x_{(Blue-NIR)/(Blue+NIR)}$ \\ 
Snow NDSI & $x_{(Blue-SWIR)/(Blue+SWIR)}$ \\
Red-SWIR ratio & $x_{Red/SWIR}$ \\
NDVI & $x_{(NIR-Red)/(NIR+Red)}$ \\
\hline
\hline
\end{tabular}
\end{center}
\end{table}

\subsection{Supervised Classification Algorithms}

The extracted features and the original spectral bands are used as inputs of advanced supervised classification algorithms, which are required to solve complex classification problems such as cloud masking. 
The detection of clouds can be considered as a two-class classification problem. In these problems, we are given a set of $\ell$ labeled (training) samples $\{\mathbf{x}_i,y_i\}_{i=1}^\ell$, where $\mathbf{x}_i\in {\mathbb R}^d$ is defined in an input space ${\mathcal X}$, and $y_i \in \{0,1\}$ belongs to the observation (output) space (`cloudy' or `cloud free'). 
In this paper, different classification methods are analyzed: classification trees (TREE) \cite{Breiman84}, support vector machines (SVMs) \cite{Scholkopf02}, and multilayer perceptron (MLP) neural networks \cite{Haykin99}. 

The {\em CART algorithm} is a tree graph structure with a sequence of nodes that are partitioned or split into two branches by means of decision rules and each terminal node (leaf) is classified with the predicted value for that node. 
Pruning and cross-validation methods are usually employed on CART algorithms to avoid overfitting. 
In this work, a $10$-fold cross-validation procedure is used to find the minimum-cost tree and to estimate the best level of pruning. 

The {\em SVM binary classifier} is a statistical learning algorithm based on constructing a maximum margin separating hyperplane in a reproducing kernel Hilbert space. 
SVMs allow the use of a high number of input features as it combats the curse of dimensionality efficiently. 
In this work, we use a Gaussian RBF kernel, given by $K({\mathbf x}_i,{\mathbf x}_j)= \exp{(- \|{\mathbf x}_i -{\mathbf x}_j\|^2 / 2\sigma^2 )}$; and free SVM parameters are selected by following an $8$-fold cross-validation procedure in the training set. 

The {\em MLP neural network} \cite{Haykin99}, which has been a traditional approach for supervised cloud classification \cite{Torres-Arriaza03}, is also included in the comparison. 
In all the cases, the neurons of the hidden layer present the hyperbolic tangent sigmoid activation function while the neuron of the output layer presents a linear output function in order to better analyze the distribution of the output values.

\section{Experiments \& Results}\label{sec:results}

\begin{figure}[t]
\begin{center}
\includegraphics[width=9cm]{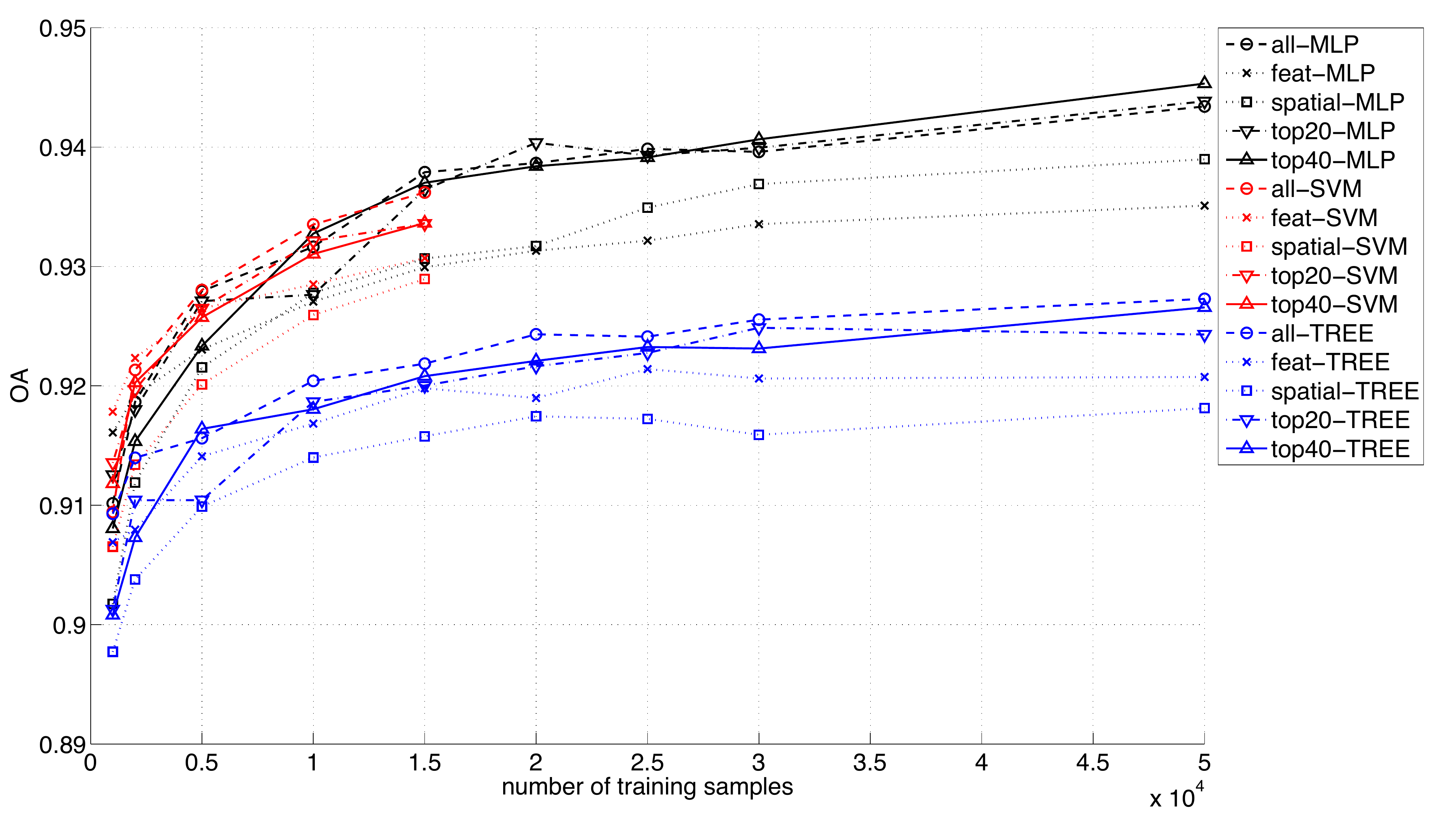}
\caption{ \small \label{fig:summary_methods} Overall Accuracy (OA\%) over the test sets for the analyzed methods (TREE, SVM and MLP). The number of input features (spectral, spatial, and all features) and training samples per class vary for each test set.}
\end{center}
\end{figure}

In order to decide which features are more relevant, the information of the spectral channels and the extracted features for cloud detection is analyzed in terms of classification accuracy.  
In the following experiments, classifiers are trained with different numbers of training samples and different combinations of features. 
Figure~\ref{fig:summary_methods} shows the Overall Accuracy (OA) for each number of features, i.e. the classification accuracy of the TREEs, MLPs and SVMs for the sets of selected features.  

Several conclusions can be extracted from Fig.~\ref{fig:summary_methods}. 
Classification trees are very efficient classification algorithms but provide the less accurate detection results for all cases. 
SVMs provide excellent results when few training samples are available but present a huge computational cost when the number of samples increases (no SVM models have been trained for more than 15000 training samples per class).  
Finally, MLP neural networks provide excellent cloud detection accuracy and the extracted spatio-spectral cloud features drastically improve results, obtaining the best results with the top 40 selected features. 

\begin{figure}[t]
\begin{center}
\includegraphics[width=7cm]{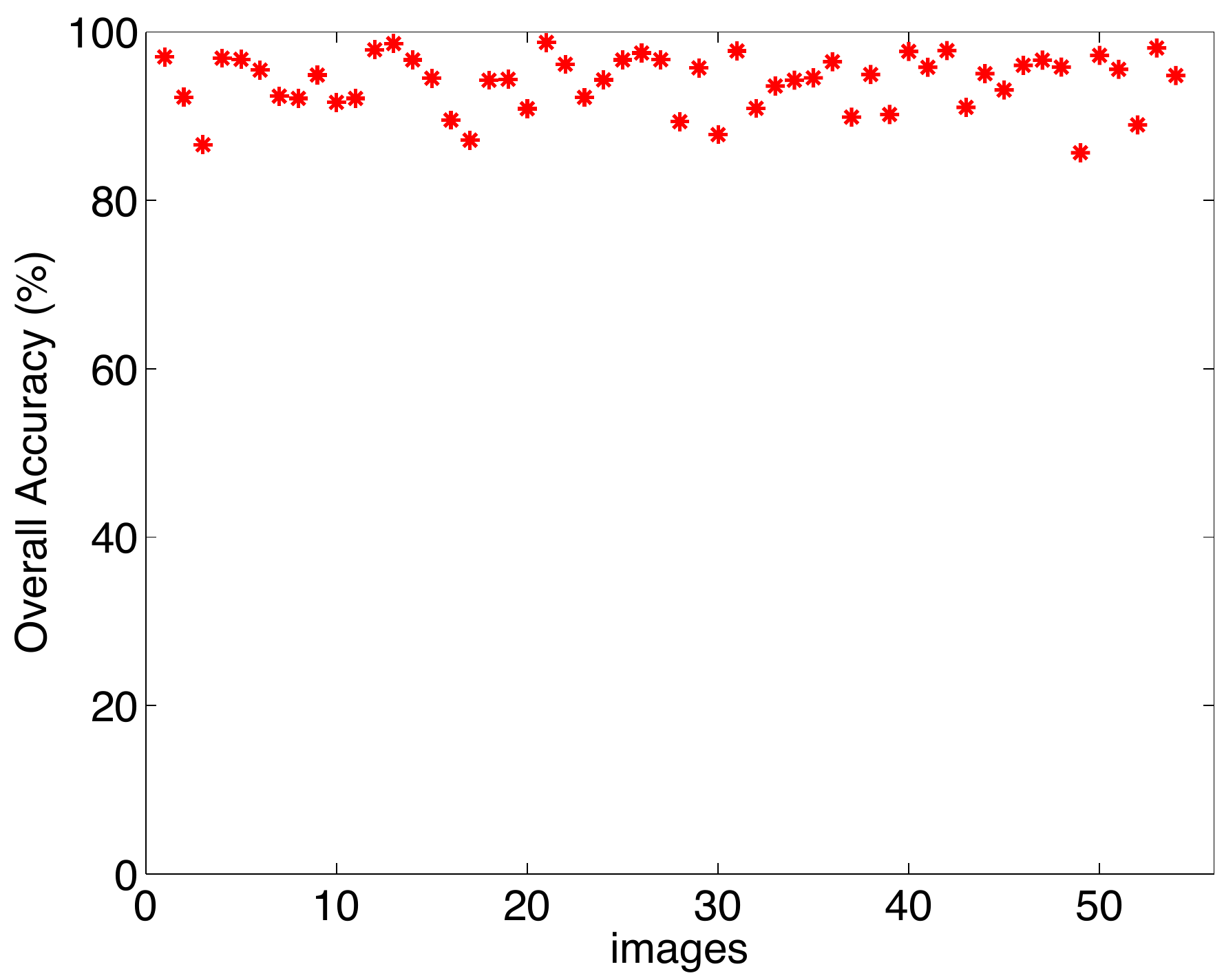}
\caption{ \small \label{fig:summary_images} Overall Accuracy (\%) 
for the 54 validation images manually labeled in order to be used as reference (ground truth). }
\end{center}
\end{figure}

In what follows, we focus only on MLPs, which have offered improved performance, using the top 40 selected features for Proba-V. Also, and given that accuracy is not improved too much with an increasing number of samples, we concentrate on using $50,000$ training samples per class in all cases, and classification accuracy is computed using $380,000$ test samples. 
Figure~\ref{fig:summary_images} shows the Overall Accuracy (OA\%) 
for the 54 validation images that have been manually labeled in order to be used as reference (ground truth). 
In this figure, one can observe that in most images the cloud detection accuracy is higher than 90\%. 
This confirms that the trained MLP provides an excellent generalization over the analyzed images. 
However, it is important to remark that we are using as reference cloud mask a `ground truth' that has been manually generated. Hence, the risk of learning and reproduce the errors present in the ground truth there always exists.   

In order to better analyze the type of errors that we are committing, Table~\ref{tab:confusion_matrix} shows the overall Confusion Matrix for the 54 validation images manually labeled. 
One can observe that the proposed method provides a balanced number of false negatives (FN) and false positives (FP), although the number of FN is relatively higher. However, the average Overall Accuracy for all images is 93\% and we can consider that the agreement between the predicted cloud masks and the generated ground truth is high enough. 

\begin{table}[t]
\caption{ \small \label{tab:confusion_matrix}
 Confusion Matrix for the 54 validation images that have been manually labeled in order to be used as reference (ground truth). True Negatives (TN), False Negatives (FN), True Positives (TP), False Positives (FP), Producer Accuracy (PA), User Accuracy (UA), and Overall Accuracy (OA). }
\begin{center}
\footnotesize
\begin{tabular}{r|cc|l}
\cline{2-3}
& \multicolumn{2}{c|}{\textbf{Manual Labels}} & \\
\cline{2-3}
& \textbf{Cloud-free} & \textbf{Cloudy} &  \\
\cline{2-3}
\textbf{Predicted Cloud-free} & \cellcolor[rgb]{.7,.7,.7} \textbf{TN}: 345$\times10^6$ & \cellcolor[rgb]{.9,.9,.9} \textbf{FN}: 25$\times10^6$ & \textbf{PA}: 93\% \\
\textbf{Predicted Cloudy} & \cellcolor[rgb]{.9,.9,.9} \textbf{FP}: 9$\times10^6$  & \cellcolor[rgb]{.7,.7,.7} \textbf{TP}: 191$\times10^6$  & \textbf{PA}: 95\% \\
\cline{2-3}
& \textbf{UA}: 97\% & \textbf{UA}: 88\% & \textbf{OA}: 93\% \\
\cline{2-3}
\end{tabular}
\end{center}
\end{table}

\begin{figure}[h]
\begin{center}
\footnotesize
\begin{tabular}{cc}
RGB (2014/09/21) & Mask Comparison \\
\includegraphics[width=4cm]{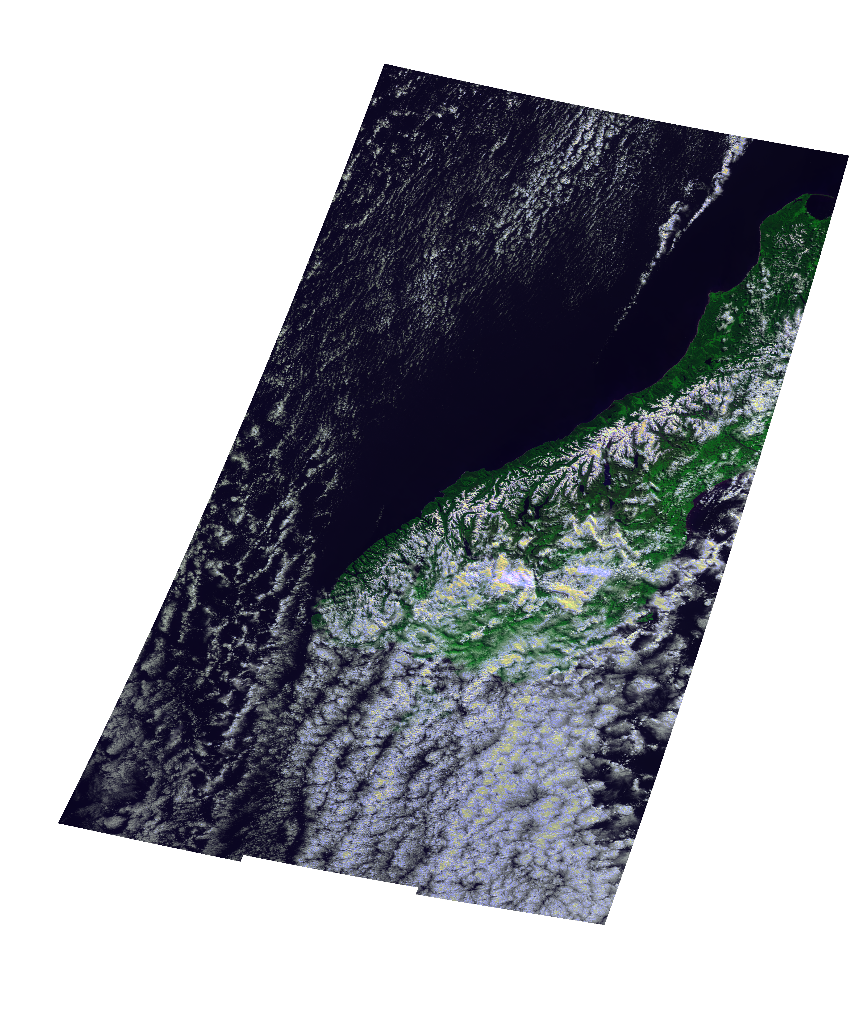}
& \includegraphics[width=4cm]{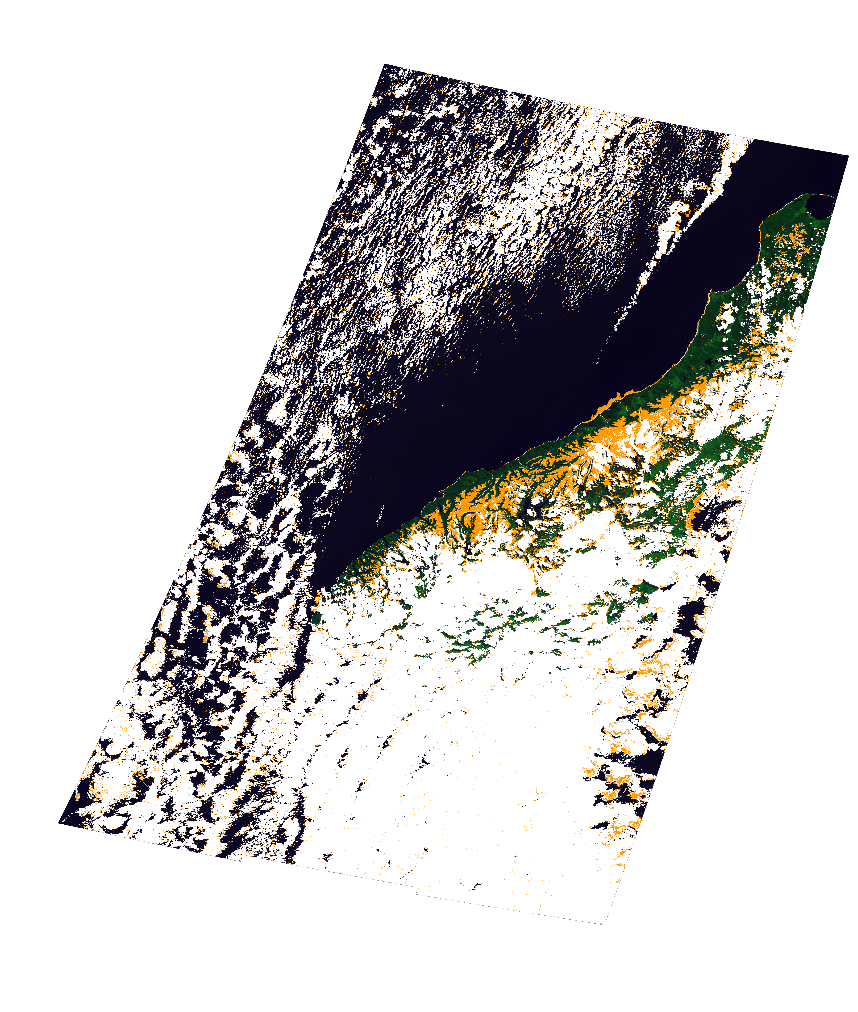}
\vspace*{-0.5cm}
\\ 
Manual Ground Truth & Predicted Cloud Mask   
\\ \includegraphics[width=4cm]{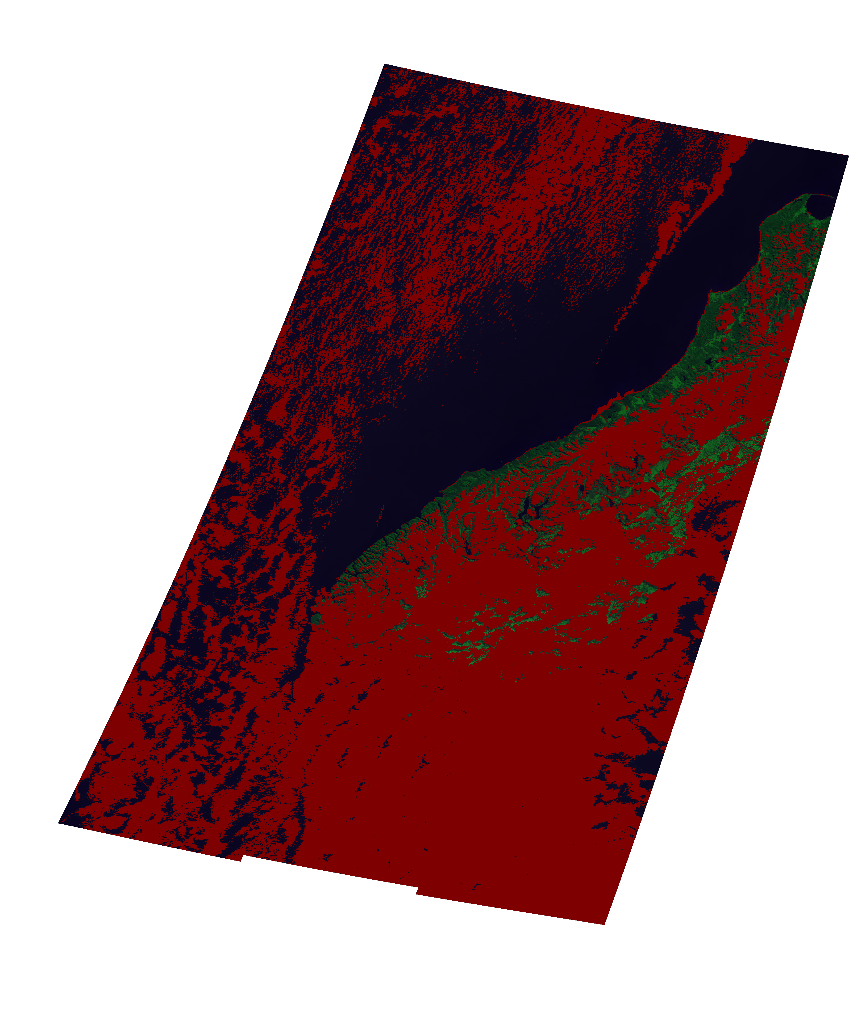}
& \includegraphics[width=4cm]{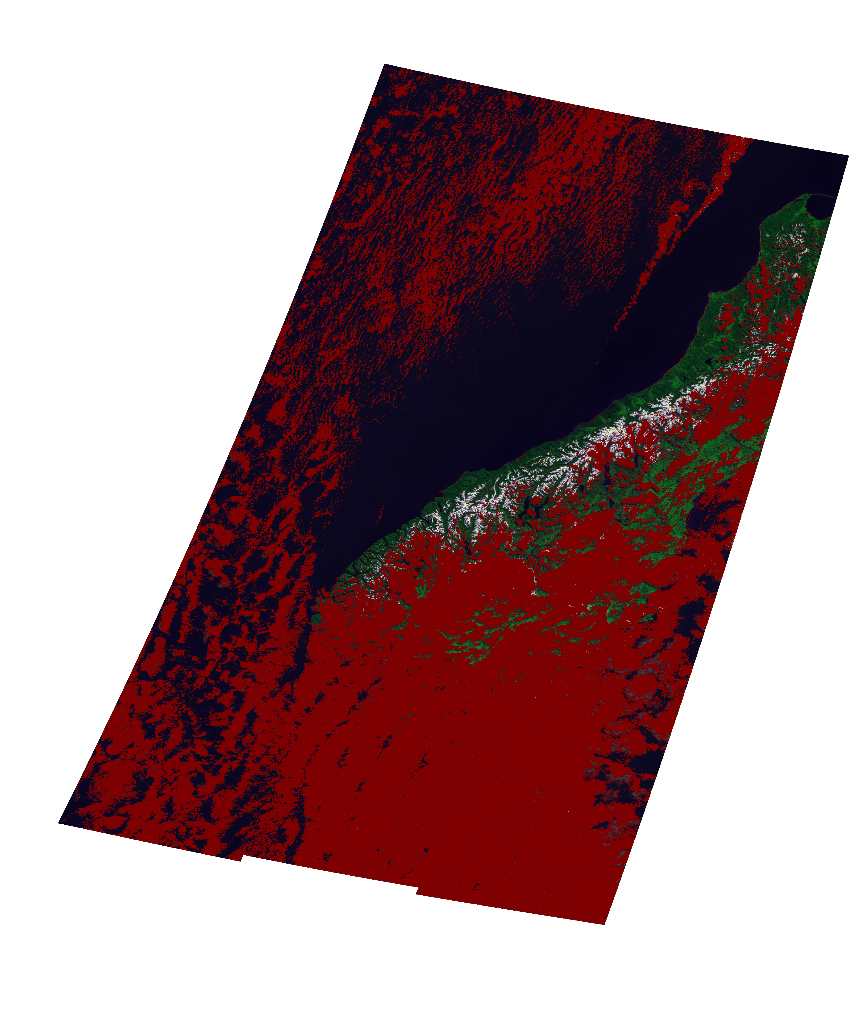}
\end{tabular}
   \scriptsize
   \begin{tabular}{p{2.2cm}p{1.1cm}p{1.0cm}p{1.0cm}p{1.0cm}}
Proposed / Reference: & \makebox[1.1cm][c]{Cloud / Cloud} & \makebox[1.1cm][c]{Land / Cloud} & \makebox[1.1cm][c]{Cloud / Land} &\makebox[1.1cm][c]{Land / Land} 
   \end{tabular}
   \begin{tabular}{p{2.2cm}|p{1.1cm}|p{1.0cm}|p{1.0cm}|p{1.0cm}|}
   \cline{2-5}
Comparison Color: & \cellcolor[rgb]{1,1,1} & \cellcolor[rgb]{1,0.5,0} & \cellcolor[rgb]{0,0,1} & \cellcolor[rgb]{0,0,0} \\
   \cline{2-5}
   \end{tabular}
   \vspace*{-0.5cm}
\end{center}
\caption{ \small \label{fig:cloudmask_example}   Cloud detection example showing the RGB false color composite, the manually generated ground truth, the cloud mask obtained with the final MLP classifier, and the comparison of the `manual ground truth' with the `Proba-V Cloud Flag': discrepancies are shown in blue when proposed method detects cloud and in orange when pixels are classified as cloud-free.}
\end{figure}

Finally, an example of the resulting cloud mask is shown in Fig.~\ref{fig:cloudmask_example}. In this figure, the proposed cloud mask is benchmarked against the manually generated `ground truth' for a Proba-V product. This image has been selected because it illustrates most common cloud detection problems (e.g. cloud borders, thin clouds, ice/snow covers). 
In these plots, pixels detected as cloudy pixels by the cloud detection method but labeled as cloud free in the `ground truth' are plotted in \emph{blue}, while discrepancy pixels classified as cloud free but marked as clouds in the `ground truth' are shown in \emph{orange}. 
The pixels detected as cloud-free by the algorithm that are labeled as cloudy in the `ground truth' (\emph{orange}) match snow and glaciers over mountains in the South Island of New Zealand, and some parts of the coastline.   
Therefore, one can assume that these differences are not errors but cloudy pixels, and that proposed method recognizes these difficult cases, although it is true that some thin cloud borders are also detected as cloud-free. However, in these cases a spatial growing of the cloud mask should easily improve the results.

\section{Conclusions}\label{sec:conclusions}

In this paper, a methodology that faces the problem of accurately identifying the location of clouds in Proba-V images is described. 
The cloud masking algorithm is based on simple spatio-spectral physical features, which are intended to increase separability between clouds and ground covers, and are extracted from the converted top of atmosphere reflectance in order to reduce dependence on illumination and geometric acquisition conditions. 
A supervised classification is carried out based on the selected extracted features and selected training samples covering most relevant image conditions, background surfaces, and cloud types. 
In particular, several machine learning methods have been trained using different sets of input features and different sets of training samples in order to select the best empirical model. The final implemented method is based on artificial neural networks trained with manually labeled real data. 
The performance of the method has been tested on a large number of real images and on scenes presenting most critical cloud detection problems, and results show an accurate discrimination of thin clouds, cloud borders, and bright surfaces.

\small
\bibliographystyle{IEEEbib}
\bibliography{}

\end{document}